\documentclass[cits]{PoS}

\usepackage{url}
\usepackage{amsmath}

\makeatletter
\newcommand{\fmslash}[2][0mu]{%
  \mathchoice
    {\fmsl@sh\displaystyle{#1}{#2}}%
    {\fmsl@sh\textstyle{#1}{#2}}%
    {\fmsl@sh\scriptstyle{#1}{#2}}%
    {\fmsl@sh\scriptscriptstyle{#1}{#2}}}
\newcommand{\fmsl@sh}[3]{%
  \m@th\ooalign{$\hfil#1\mkern#2/\hfil$\crcr$#1#3$}}
\makeatother

\def\bra#1{\mathinner{\langle{#1}|}}
\def\ket#1{\mathinner{|{#1}\rangle}}
\def\braket#1{\mathinner{\langle{#1}\rangle}}
\newcommand{\ii}{\mathrm{i}}

\title{Recent developments for multi-leg QCD amplitudes with massive particles}
 
\ShortTitle{Recent developments for multi-leg QCD amplitudes with massive particles}

\author{ 
Rutger Boels\\
 Niels Bohr Institute,
 Niels Bohr International Academy\\
Blegdamsvej 17, DK-2100 Copenhagen, Denmark\\
and\\
 The Mathematical Institute, University of Oxford\\
24-29 St. Giles, Oxford OX1 3LP, United Kingdom
}

\author{\speaker{Christian Schwinn} %
\thanks{Supported by the  DFG Sonder\-forschungsbereich/Transregio~9
``Computergest\"utzte Theoretische Teilchenphysik''}\\
Institut f\"ur Theoretische Physik E\\
RWTH Aachen, D - 52056 Aachen, Germany\\
E-mail: \email{schwinn@physik.rwth-aachen.de}
}

\author{Stefan Weinzierl\\
       Institut f{\"u}r Physik, Universit{\"a}t Mainz\\
      D - 55099 Mainz, Germany}  

 \abstract { 
        
We review the extension of modern techniques for the calculation of helicity amplitudes in QCD to massive particles. The focus is on the use of supersymmetric Ward identites that relate amplitudes with massive quarks to those with
massive scalars, the application of on-shell recursion relations to amplitudes with massive quarks and an extension of the CSW rules to massive scalars.

}
 \dedicated{
December 20, 2007\\
PITHA~07/24 \\
SFB/CPP-07-92\\
}
\FullConference{8th International Symposium on Radiative Corrections \\
		 October 1-5, 2007\\
		 Florence, Italy}

\begin{document}

\section{Introduction}
Witten's observations on twistor-space properties of QCD scattering
amplitudes~\cite{Witten:2003nn} led to a number of new calculational
methods like the diagrammatic rules of Cachazo, Svr\v{c}eck and
Witten~(CSW) \cite{Cachazo:2004kj}, the on-shell recursion relations
of Britto, Cachazo, Feng and Witten~(BCWF)
\cite{Britto:2004ap,Britto:2005fq} and a renewed interest in unitarity
methods for loop calculations~\cite{Bern:2007dw}.  Several challenges
remain to turn these methods into viable alternatives to traditional
approaches e.g. automatization, the numerical stability and the
extension to massive particles.  This contribution reviews progress in
the latter area. In section~\ref{sec:susy} Supersymmetry~(SUSY) is
used to relate the simplest helicity amplitudes of massive quarks and
scalars.  Extensions of the BCFW and CSW rules to massive quarks and
scalars are presented in sections~\ref{sec:bcfw} and~\ref{sec:csw}.

\section{MHV amplitudes and SUSY Ward-Identities for massive quarks and scalars}\label{sec:susy}
The simplest helicity amplitudes of gluons are the "maximally helicity
violating"~(MHV) amplitudes with two negative helicity gluons and an
arbitrary number of positive helicity gluons~\cite{Parke:1986gb}. In
spinor-helicity notation the color-ordered partial amplitudes (see
e.g.~\cite{Bern:2007dw}) read
\begin{equation}
\label{eq:mhv}
  A_n(g_1^+,\dots,g_i^-,\dots, g_j^-,\dots g_n^+)
=\ii 2^{n/2-1}
\frac{\braket{ij}^4}{\braket{12}\braket{23}\dots\braket{(n-1)n}\braket{n1}}.
\end{equation}
Color-stripped tree amplitudes of massless quarks in QCD are identical
to those of gluinos $ \Lambda$ in an unbroken SUSY Yang-Mills theory.
Using the fact that the SUSY charge 
annihilates the vacuum, one can derive SUSY Ward-identities~(SWIs) in
an unbroken SUSY Yang-Mills theory:
\begin{equation}
\label{eq:swi}
  0=\braket{0|\lbrack Q_{\text{SUSY}}(\eta), \Phi_1\dots \Phi_n\rbrack|0}
=\sum _i\braket{0|\Phi_1 \dots \delta_{\eta} \Phi_i  \dots \Phi_n|0}.
\end{equation}
 The SUSY transformations of the helicity
states parameterized by an anti-commuting spinor
$\eta$ read
\begin{equation}
\label{eq:susy-gluon}
 \delta_\eta g^\pm 
 = \Gamma_\eta^\pm(k)  \Lambda^\pm\;,\quad
 \delta_\eta \Lambda^\pm =\Gamma^\mp_\eta(k) g^\pm
\end{equation} 
where $\Gamma^\pm_\eta(k)= \sqrt 2 \braket{\eta\pm|k\mp}$. 
These transformations can be used in the SWI to relate 
the MHV amplitudes~\eqref{eq:mhv}
to amplitudes with massless quarks~\cite{Parke:1985pn}:
\begin{equation}
\braket{1j}A_n(\bar \Lambda_1^-,g_2^+,\dots ,g_j^-,\dots,\Lambda_n^+)
=\braket{nj}
A_n(g_1^-,g_2^+,\dots, g_j^-, \dots,g_n^+).
\end{equation}

The MHV amplitudes can serve as building blocks for
\emph{all} massless QCD amplitudes, either as vertices in CSW
diagrams~\cite{Cachazo:2004kj} or as input in 
BCFW relations~\cite{Britto:2004ap,Britto:2005fq}.
As a  starting point for the extension of such methods one can
therefore consider the simplest amplitudes with massive particles, 
e.g. amplitudes with a pair of massive scalars 
and positive helicity gluons~\cite{Ferrario:2006np}
\begin{equation}
\label{eq:rodrigo}
  A(\bar \phi_1,g_2^+,\dots,\phi_n)
    = \ii 2^{n/2-1}m^2 \frac{  \braket{2+|\prod_{j=3}^{n-2}
\left(y_{1,j}-\fmslash k_j\fmslash k_{1,j}\right)|n-1} }{
y_{1,2} y_{1,3}\dots y_{1,n-2} 
\braket{23}\braket{34}\dots\braket{(n-2)(n-1)}} 
\end{equation}
where $k_{i,j}= k_i+\dots k_j$ and $ y_{1,j}=k_{1,j}^2-m^2$.
The amplitudes~\eqref{eq:rodrigo} can be  related to amplitudes with massive quarks using SWIs in a  SUSY 
Yang-Mills theory with a massive quark $Q$ 
and two complex massive scalars $\phi^\pm$
as super-partners~\cite{Schwinn:2006ca}. 
External states of massive quarks can be introduced
as  
\begin{eqnarray}
\label{eq:spinors} 
 u(\pm) = \frac{1}{\braket{p^\flat\mp| q\pm}}
 \left( \fmslash p + m \right) \ket{q \pm} ,
 & &
\bar{u}(\pm) = \frac{1}{\braket{ q\mp|  p^\flat\pm}} \bra{ q \mp}
 \left( \fmslash p + m \right)
\end{eqnarray}
with the light-like vector $ p^\flat = p - p^2/ (2 p \cdot q) q$ and
where $q$ defines the axis of the quark spin.  The ``helicity'' states
of the quarks defined by the spinors~\eqref{eq:spinors} are related to
the scalars by SUSY transformations
  \begin{equation}
\delta_\eta Q^\pm=-\Gamma^\pm_\eta(k) \phi^\pm + \Sigma^\mp_\eta(k,q) \phi^\mp
\;,\quad
 \delta_\eta \phi^\pm = -\Gamma^\mp_\eta(k) Q^\pm - \Sigma^\mp_\eta(k,q) Q^\mp 
  \end{equation}
where $\Sigma^\pm_\eta(k,q)=\sqrt 2 
  m\braket{q\pm|\eta\mp}/\braket{q\pm|k\mp}$.
For the choice $\ket{\eta+}\propto \ket{q+}$ the terms proportional to the mass
drop out and the transformations become
similar to the massless case~\eqref{eq:susy-gluon}.
The amplitudes with massive quarks and
positive helicity gluons are related to the scalar amplitudes~\eqref{eq:rodrigo} by a simple
SWI:
\begin{equation}
\braket{1q}  A(\bar Q_1^+,g_2^+,\dots,Q_n^-)=\braket{nq}
    A(\bar \phi_1^+,g_2^+,\dots,\phi_n^-).
\end{equation}
The amplitudes with one negative helicity gluon can also 
be related 
to scalar amplitudes~\cite{Schwinn:2006ca}.

\section{On-shell recursion relations for massive quarks}
\label{sec:bcfw}
The on-shell recursion relations~\cite{Britto:2004ap,Britto:2005fq,Badger:2005zh} 
express tree amplitudes
 in terms of  products
 of two amplitudes with fewer external particles:
\begin{equation}
\label{eq:bcfw}
        A_n(\Phi_1,\dots,\Phi_n)=\!\!\!\!
         \sum_{P(i,j),\sigma }^{n}\!\!\!\!\!\!
  A(\Phi_r, \dots,\Phi_i',\dots,\Phi_s, - {\Phi_K'}^\sigma )         
       \frac{\ii}{K^2-m^2}A({\Phi_K'}^{-\sigma} ,
\dots, \Phi_j' ,\dots , \Phi_{r-1}).
\end{equation}
The sum is over 
helicities $\sigma$ and 
all partitions of momenta $P(i,j)$ into two sets so that $k_i$ is in
one of the sets and $k_j$ in the other one.   
On the right hand side
the two external momenta $k_i$ and $k_j$ 
are shifted into the complex plane.
For light-like $k_i$ and $k_j$ 
the shift is performed by shifting the spinors according to
\begin{equation}
\label{eq:shift}
\ket{i'+}=\ket{i+}-z\ket{j+}\;,\qquad 
\ket{j'-}=\ket{j-}+z\ket{i-}.
\end{equation}
In each term of the relation~\eqref{eq:bcfw}, 
the value of $z$ is chosen so that the internal momentum
$K'=k_r+\dots+k_i'+\dots+ k_s$ is on-shell.

One approach for the application of on-shell recursion relations to
amplitudes of massive particles with spin~\cite{Badger:2005jv}
expresses \eqref{eq:bcfw} in terms of ``stripped'' amplitudes with
removed external polarization spinors.  Recently all amplitudes with a
pair of massive quarks and up to four gluons have been calculated in
this framework~\cite{Ozeren:2006ft}. As shown in~\cite{Schwinn:2007ee}
one can also treat internal quarks in the helicity formalism which
allows to shift also massive quark lines. This can be used to show
that \emph{all} QCD amplitudes (i.e. also those with only massive
quarks) can be obtained from BCFW relations and simplifies the
structure of the recursion relations in some cases.  The shift of two
massive momenta $p_{i/j}$ is implemented by decomposing them into two
two light-like vectors $l_{i/j}$
\begin{equation}
p_i = l_i + \alpha_j l_j \;,\quad
p_j = \alpha_i l_i + l_j.
\end{equation}
For explicit expressions see~\cite{Schwinn:2007ee}.
The shift of massive-quark spinors 
$u_i(-)$ and $\bar{u}_j(+)$ is defined as
\begin{eqnarray}
\label{eq:massive-shift}
 {u_i}'(-) = u_i(-) - z \ket{ l_j + },
 & &
 {\bar{u}_j}'(+) = \bar{u}_j(+) + z \bra{l_i + }.
\end{eqnarray}
Here $\ket{l_j\pm }$ have to be used as reference
spinors for a massive quark  $Q_i$ 
(and vice versa for $Q_j$) in order to obtain completeness relations 
and to avoid spurious poles in $z$~\cite{Schwinn:2007ee}.
For an internal massive quark  $Q_K$ in~\eqref{eq:bcfw}, one  uses the
reference spinors $\ket{l_j +}$ and $\bra{ l_i +}$.

The proof of the BCFW relations in~\cite{Britto:2005fq} continues the
scattering amplitude to arbitrary complex values of $z$ and requires
$\lim_{z\to \infty} A(z)=0$. This constrains the helicity of the
particles $i$ and $j$.  By estimating the $z\to \infty$ behavior of
individual Feynman diagrams as in~\cite{Britto:2005fq} it follows that
the case $(i^+, j^-)$ is always allowed unless $i$ and $j$ are quarks
joined by a fermion line~\cite{Luo:2005rx}. Using a supplementary
three particle shift~\cite{Badger:2005zh,Schwinn:2007ee} one can show
that also the combinations $(g_i^+,g_j^+)$, $(g_i^+, Q_j^+)$ and
$(q_i^+,Q_j^+)$ for a massless quark $q_i$ are allowed.  Amplitudes
with only massive quarks can be obtained by shifting three external
legs. The case $(i^-, j^-)$ is analogous~\cite{Schwinn:2007ee}.

\section{CSW diagrams for massive scalars}
\label{sec:csw}
In the CSW rules~\cite{Cachazo:2004kj} QCD amplitudes are constructed
from diagrams with vertices that are given by off-shell continuations
of the MHV amplitudes~\eqref{eq:mhv}.  External massive gauge- and
Higgs bosons have been included in this approach
in~\cite{Dixon:2004za,Bern:2004ba}.  In~\cite{Boels-CS} it is shown
how to extend the CSW rules to propagating massive scalars using a
canonical transformation method~\cite{Mansfield:2005yd} and by the
construction of a twistor action that reduces to the CSW-Lagrangian in
a certain gauge~\cite{Boels:2007qn}.  It is expected that similar
rules for massive particles with spin can be derived in an analogous
way.

In the approach of~\cite{Mansfield:2005yd}, one works with the
light-cone gauge Lagrangian that contains only the physical modes of
the gluon, $A_z$ (positive helicity) and $A_{\bar z}$ (negative
helicity), with interactions of the helicity structure
$\mathcal{L}^{(3)}_{++-}$, $\mathcal{L}^{(3)}_{+--} $ and
$\mathcal{L}^{(4)}_{++--} $.  A canonical transformation to new
variables $B[A_z]$ and $\bar B[A_z,A_{\bar
  z}]$~\cite{Mansfield:2005yd} can be used to eliminate the non-MHV
vertex $\mathcal{L}^{(3)}_{++-}$ in favor of a tower of MHV-vertices $
\mathcal{L}^{(n)}_{+\dots +--}$.  A similar procedure can be applied
to the light-cone gauge Lagrangian for scalars with the structure
\begin{equation}
 \mathcal{L}^{(2)}(\bar\phi\phi)+ 
\mathcal{L}^{(3)}(\bar\phi A_z\phi)
+\mathcal{L}^{(3)}(\bar\phi A_{\bar z}\phi)
+\mathcal{L}^{(4)}(\bar\phi A_zA_{\bar z}\phi)
+\mathcal{L}^{(4)}(\bar\phi \phi\bar\phi\phi)
\end{equation}

Using the \emph{same} transformation 
$  \phi\to \xi[\phi,A_z]$ for massive and massless scalars
in addition to the transformation of the gluons,
one can eliminate  the
non-MHV type coupling 
 $\mathcal{L}^{(3)}(\bar\phi A_z\phi)$ and obtains
the MHV vertices for massless
scalars~\cite{Georgiou:2004wu} such as
\begin{equation}
\label{eq:csw-2phi}
  V_{\text{CSW}}(\bar \xi_1,g^+_2,\dots g_{i}^-,\dots \xi_n)=
-\ii 2^{n/2-1}  \frac{ \braket{in}^2\braket{1i}^2}{
  \braket{12}\dots\braket{(n-1)n}\braket{n1}}
\end{equation}
and an additional tower
of vertices with a pair of scalars and an arbitrary
number of positive helicity gluons that is
generated from the transformation of the mass term~\cite{Boels-CS}:
 \begin{equation}
\label{eq:csw-mass}
V_{\text{CSW}}(\bar \xi_1,g^+_2,\dots \xi_n)= \ii  2^{n/2-1}
\frac{- m^2 \braket{1n}}{
\braket{12}\dots\braket{(n-1)n }} 
\end{equation}

As an application, consider the proof of the BCFW recursion relation for
a $(g_i^+,g_j^+)$ shift. For gluon amplitudes this follows from the CSW
representation~\cite{Britto:2005fq}. The rules given here 
allow to extend this argument  to amplitudes with massive scalars~\cite{Boels-CS}, without recourse to auxilary shifts.

\section{Conclusions}
We have reviewed the extensions of new methods for the calculation of
scattering amplitudes to massive particles.  While we have focused on
results for QCD tree amplitudes including massive quarks or scalars,
first steps for the extension to the electro-weak theory have
appeared~\cite{Bern:2004ba,Badger:2005jv} and some of the on-shell
approaches to loop calculations~\cite{Bern:2007dw} generalize to the
massive case.

\providecommand{\href}[2]{#2}\begingroup\raggedright
\endgroup

\end{document}